Submitted on June 25, 2026

# Spin-Orbit-Torque-Driven Recurrent Motion of Synthetic Antiferromagnetic Skyrmions for Magnetic Clocking

Sayan Mathur[1,†], Payal Bhattacharjee[1,†], Saswati Barman[2,3] and Anjan Barman[1,*]

[1]Department of Condensed Matter and Materials Physics, S. N. Bose National Centre for Basic Sciences, Block JD, Sector III, Salt Lake, Kolkata 700106, India

[2]Department of Basic Science and Humanities, Institute of Engineering & Management, Salt Lake Electronics Complex, Sector V, Salt Lake, Kolkata 700091, India

[3]University of Engineering and Management, University Area, Plot No. III, B/5, New Town Road, Action Area III, Newtown, Kolkata 700160, India

[†]Authors with equal contribution

[*]Author(s) to whom correspondence may be addressed: abarman@bose.res.in

**Abstract:** Magnetic skyrmions in synthetic antiferromagnets have emerged as attractive information carriers for spintronic technologies due to their compensated topology, which mitigates the skyrmion Hall effect and facilitates reliable current-driven transport. Here we present a frequency-tunable magnetic clocking concept based on the recurring translational motion of SAF skyrmions and define the material conditions needed to realize it in a Pt/Co/Ru/Pt/Co multilayer. Using micromagnetic simulations, we investigate the combined effects of interfacial Dzyaloshinskii-Moriya interaction and interlayer Ruderman-Kittel-Kasuya-Yosida exchange coupling, delineating the parameter space in which coupled Néel-type skyrmions with opposite topological charges form a stable, topologically compensated pair. Under spin-orbit torque excitation, the SAF skyrmions exhibit resilient longitudinal transport with minimal transverse deflection using periodic boundary conditions. Periodic boundary conditions serve as an idealized model for closed-loop racetrack geometries, facilitating the study of continuous transport cycles and long-range steady-state dynamics. The oscillation frequency is readily modified by the applied current density and material characteristics, offering a pathway to field-free nanoscale timing components. Moreover, the threshold-activated initiation of recurrent transport mimics tonic-spiking behavior, indicating possible use in neuromorphic signal generation. This work establishes SAF skyrmions as versatile building elements for future spintronic functional devices by integrating topological protection, controllable transport, and cyclic dynamics.

## Introduction

Due to their significance and applications in next-generation spintronics, magnetic topological spin textures, including magnetic skyrmions, have ignited the interest of researchers.[1-3] Magnetic skyrmions are topologically nontrivial chiral spin patterns characterized by whirling configurations. The topological charge or the skyrmion number ($Q = \frac{1}{4\pi}\int \boldsymbol{m} \cdot (\frac{\partial \boldsymbol{m}}{\partial x} \times \frac{\partial \boldsymbol{m}}{\partial y}) dxdy$) is ±1, indicating that the skyrmion magnetization encircles the spin unit sphere once. Their topological stability, along with their minuscule lateral dimensions down to the nanoscale limit, assigns particle-like characteristics to them.[3] Skyrmion-based devices demonstrate potential for rapid response, non-volatile, ultra-low power logic and memory systems. Topologically protected skyrmions are nanoscale spin-based quasiparticles that can be created, annihilated, and manipulated by applying an electric current or a magnetic field.[4-6] A significant first step toward the realization of such devices was the recent development of room-temperature skyrmions in technology-relevant ultra-thin ferromagnetic (FM) films lacking inversion symmetry and their simultaneous manipulation by current.[7-9] The crucial importance of the interfacial Dzyaloshinskii–Moriya interaction (iDMI) in the nucleation and stability of skyrmions has drawn a lot of interest to the study of chiral magnetic interactions. Recent studies have shown that the Heisenberg exchange interaction and the iDMI have a critical relationship, and these interactions enable the nucleation and stability of magnetic skyrmions in various systems.[10, 11, 12]

However, in traditional ferromagnetic systems, skyrmions exhibit a transverse deviation from the direction of the applied driving force, a phenomenon known as the skyrmion Hall effect (SkHE).[7] This deflection occurs because of the non-trivial topology of the skyrmion, which generates a Magnus force ($G\boldsymbol{z}\times\boldsymbol{v}$; G represents the amplitude of the skyrmion gyrovector, which is proportional to the skyrmion topological charge (Q) and the layer magnetic moment; $\boldsymbol{v}$ denotes the skyrmion velocity vector, and $\boldsymbol{z}$ is the unit vector perpendicular to the film plane) perpendicular to the direction of motion. The magnitude of this transverse motion varies and is influenced by multiple factors, including the driving velocity, the size and shape of the skyrmion, and the properties of the underlying pinning landscape formed by defects, edge roughness, or grain boundaries.[13,14] Consequently, when a skyrmion advances along a nanotrack, it tends to deviate towards one side of the track instead of adhering to a purely longitudinal path. This lateral drift significantly increases the probability of undesired annihilation at device edges, presenting a substantial obstacle for the practical applications of skyrmion-based technologies. The reliability and scalability of skyrmion-based memory and logic architectures are limited by the inability to prevent or control this transverse motion, especially in narrow or patterned geometries where edge proximity is inevitable.[15]

To counteract this effect, various design strategies have been put forth. These encompass engineering anisotropy,[16-18] the use of ferrimagnets,[19] antiferromagnets,[20-22] and synthetic antiferromagnets (SAF).[23-26] The latter comprises two ferromagnetic (FM) layers interspersed by a non-magnetic (NM) layer, which induces an antiferromagnetic Ruderman–Kittel–Kasuya–Yoshida (RKKY) coupling between the FM layers.[27] In SAFs, a pair of skyrmions with opposite topological charges is strongly coupled, resulting in an overall Q = 0 and hence a null SkHE. In addition, prior research has shown that SAFs devoid of the SkHE have skyrmion velocities that are equivalent to domain walls (DWs).[23,26,28] It has recently been shown by de Souza Silva *et al*.[29] that nonreciprocal gyrotropic forces originating from the internal oscillations of bound skyrmion pairs in SAFs may generate emergent self-propulsion at ultrahigh speeds. Remarkably, this propulsion spontaneously reverses due to thermal changes, which is like how bacteria move. These results demonstrate the diversity of SAF skyrmion dynamics and their potential for bio-inspired spintronic applications and unconventional transport. Although linear racetracks are typically the presumption in racetrack memory applications, the development of skyrmion-based oscillators or clock elements is a critical frontier in skyrmionics. These elements have the potential to generate tunable microwave-frequency signals that can be used in spin logic, neuromorphic computing, and communication devices.[5,30,31] Prior studies on SAF and AFM skyrmions predominantly emphasize mitigating the skyrmion Hall effect in linear transport[23–26] or investigating oscillations from internal modes, like breathing or gyrotropic motion.[30, 32–35] Conversely, the dynamics in this framework arise from the pure translational motion of SAF skyrmions within a planar configuration, facilitated by topological charge compensation and driven by direct current spin-orbit torque (DC SOT). This technique substantially differs from previous skyrmion oscillators and facilitates field-free, timing applications based on closed-loop SAF skyrmion transport.

Despite the robust current-driven transport and suppression of the skyrmion Hall effect, the potential of SAF skyrmions as dynamic timing elements remains largely unexplored. We propose a frequency-tunable magnetic clocking mechanism based on the recurring translational motion of SAF skyrmions induced by spin–orbit torque. Through micromagnetic simulations and analytical modelling, we investigate skyrmion formation in a Pt/Co/Ru/Pt/Co multilayer and determine the range of interfacial Dzyaloshinskii–Moriya interaction and interlayer exchange coupling strengths that stabilize coupled Néel-type skyrmions with opposite topological charges (Q = ±1). Under spin–orbit torque excitation, the topologically compensated skyrmion pair demonstrates steady longitudinal motion with modest transverse deflection. Utilizing periodic boundary conditions (PBCs) as an idealized model of closed-loop racetrack geometry, we illustrate continuous recurrent motion, with each full circuit representing a clock cycle. The operating frequency may be adjusted by material properties and current density, enabling a field-free nanoscale timing mechanism. In contrast to other reported skyrmion oscillators that use breathing or gyration modes, the suggested dynamics stem from stable translational recurrence facilitated by topological charge compensation. The consistently repeatable transport properties indicate possible applications in neuromorphic signal generation. These findings confirm recurrent SAF skyrmion dynamics as a viable foundation for low-power timing, oscillators, memory, and neuromorphic spintronic devices.

## Micromagnetic simulations

### A. Micromagnetic Framework: Skyrmion Formation and SOT-driven Motion in SAFs

A wide range of micromagnetic energy contributions, including exchange, anisotropy, iDMI, dipolar interaction, and, in the case of SAF, interlayer exchange coupling, interact to nucleate, stabilize, and govern the dynamics of skyrmions in ultrathin magnetic multilayers. The following represents the overall energy density of a single ferromagnetic layer:[36]

$$\varepsilon_{tot} = \varepsilon_{ex} + \varepsilon_{ani} + \varepsilon_{DMI} + \varepsilon_{dip} + \varepsilon_{ext} \tag{1}$$

where $\varepsilon_{ex}$ is the exchange energy, $\varepsilon_{ani}$ is the perpendicular magnetic anisotropy energy, $\varepsilon_{DMI}$ is the iDMI energy, $\varepsilon_{dip}$ is the dipolar (demagnetization) energy and $\varepsilon_{ext}$ is the Zeeman energy (in case an external applied field exists).

The interlayer exchange coupling between the top and bottom ferromagnetic layers, which is usually mediated by a nonmagnetic spacer such as Ru, requires the inclusion of an extra term in SAF systems, which is modeled by:[37, 38]

$$\varepsilon_{RKKY} = \frac{\sigma_{RKKY}}{t_{FM}}(1 - \boldsymbol{m}_t \cdot \boldsymbol{m}_b) \quad (2)$$

Here, $\sigma_{RKKY}$ stands for the effective interlayer exchange energy per unit area, which is considered to be negative throughout this study (unless otherwise stated), suggesting an antiferromagnetic coupling, and is dependent on the thickness $t_{FM}$ of the ferromagnetic layer. $\sigma_{RKKY}$ implicitly depends on the thickness of the non-magnetic spacer layer and does not explicitly appear within the simulations. Additionally, the unit vectors of magnetization for the top and bottom layers are denoted by $\boldsymbol{m}_t$ and $\boldsymbol{m}_b$, respectively. For a ferromagnetic arrangement, $\theta = 0^\circ$ and $\varepsilon_{RKKY} = \frac{\sigma_{RKKY}}{t_{FM}}(1 - m_t m_b)$ while for an antiferromagnetic arrangement, $\theta = 180^\circ$ and $\varepsilon_{RKKY} = \frac{\sigma_{RKKY}}{t_{FM}}(1 + m_t m_b)$. Here, $\theta$ is the polar angle of the local magnetization vector perpendicular to the film. This term energetically promotes antiparallel alignment over all spatial points, thereby imposing a skyrmion pair with opposite core polarities and opposite topological charges (Q = ±1). The dynamics of SAF skyrmions driven by SOT are characterized by the Thiele equation for a rigid skyrmion: [39,40]

$$\boldsymbol{G} \times \boldsymbol{v} - \alpha \mathcal{D} \boldsymbol{v} + \boldsymbol{F}_{SOT} = 0 \quad (3)$$

where $\boldsymbol{G}$ is the gyrovector, $\mathcal{D}$ is the dissipative tensor, $\boldsymbol{v}$ is the core velocity of the skyrmion, $\alpha$ is the Gilbert damping parameter, and $\boldsymbol{F}_{SOT}$ is the effective force due to the SOT. In SAFs, opposite skyrmion charges cancel the gyrovector contributions, yielding $\boldsymbol{G}_{total} = 0$ $(\alpha D \boldsymbol{v} = \boldsymbol{F}_{SOT})$ and hence eliminating the net SkHE. This results in a straightforward, longitudinal motion solely influenced by the damping and the applied current, without any transverse deflection. The angle between the direction of skyrmion motion and the applied current, known as skyrmion Hall angle (SkHA), is defined as:[41]

$$\tan \theta_{SkHA} = \frac{v_y}{v_x} = \frac{G}{\alpha \mathcal{D}} \quad (4)$$

This leads to a finite SkHA in ferromagnetic systems because of the non-zero topological charge, which causes the skyrmion to diverge from the current direction and often annihilate at device edges. In SAF multilayers, skyrmions in the upper and lower layers exhibit equal and opposing topological charges, resulting in a net-zero gyrovector. Consequently, the SkHE is minimized, allowing the SAF skyrmion pair to traverse only in the direction of the current without lateral deviation ($\tan \theta_{SkHA} \approx 0°$). The elimination of the SkHA is essential for attaining stable, linear motion on narrow nanotracks, allowing SAF skyrmions to function as reliable carriers in racetrack memory, logic, and oscillator devices.

### B. Geometry, Model, and Simulation Methods

An open-source GPU-accelerated micromagnetic simulator, mumax$^3$,[42,43] is employed to compute the magnetization dynamics of magnetic skyrmion based on the Landau–Lifshitz–Gilbert (LLG) equation.[44,45] Utilizing the finite-difference method, it solves the LLG equation numerically to yield micromagnetic simulation results. The magnetization dynamics on picosecond and nanosecond timescales are governed by the LLG equation, expressed as follows:

$$\dot{\boldsymbol{m}} = -|\gamma|\boldsymbol{m} \times H_{eff} + \alpha \boldsymbol{m} \times \dot{\boldsymbol{m}} + \boldsymbol{\tau}_{\mathrm{SOT}} \tag{5}$$

where $\boldsymbol{m}$ is the unit magnetization vector, $\boldsymbol{m} = \frac{\boldsymbol{M}}{M_S}$, $M_s$ is the saturation magnetization, $\boldsymbol{H_{eff}} = -\frac{1}{\mu_0}\frac{\partial \boldsymbol{E}}{\partial \boldsymbol{M}}$ is the effective magnetic field, $\gamma$ is the gyromagnetic ratio, $\alpha$ is the Gilbert damping constant, $E$ is the volume-averaged energy density, which includes magnetic anisotropy energy, the exchange energy, DMI energy, Zeeman energy, and the magnetostatic energy. The SOT term $\boldsymbol{\tau}_{\mathrm{SOT}}$ is expressed by the equation:

$$\boldsymbol{\tau}_{\mathrm{SOT}} = \frac{\beta\theta_{\mathrm{SH}}}{2(1+\alpha^2)}(\boldsymbol{m} \times (\boldsymbol{p} \times \boldsymbol{m}) - \alpha\,(\boldsymbol{p} \times \boldsymbol{m}) \tag{6}$$

where $\theta_{\mathrm{SH}}$ is the spin-Hall angle and $\beta = \frac{J_{HM}\hbar}{M_s e t_{FM}}$, $J_{HM}$ is the applied current density, $t_{\mathrm{FM}}$ is the thickness of the FM layers, $\hbar$ is the reduced Planck constant, and $e$ is the charge of an electron, $\boldsymbol{p} = -\hat{\boldsymbol{y}}$ is the direction of spin polarization induced by current injection along $+\hat{\boldsymbol{x}}$. A uniform spin polarization and current density are implemented throughout the simulation volume, and the interface-specific spin Hall injection from the upper and lower heavy-metal layers is not explicitly distinguished.

Here, we investigate a planar nanostrip of Pt/Co/Ru/Pt/Co multilayer with dimensions of 1000 nm for length, 100 nm for width, 1 nm for the thickness of the FM layers (Co layers), and 1 nm for the thickness of the spacer layer (Ru layer). However, instead of being described as explicit magnetic layers, mumax$^3$ treats heavy-metal layers (Pt, here) implicitly by integrating the interfacial phenomena they generate at the ferromagnet/heavy-metal interface, such as SOT, iDMI, etc. The system under study is discretized into unit cells with dimensions $1 \times 1 \times 1$ nm$^3$ as shown in Figure 1(a). Two magnetic skyrmions with opposite topological charge can be initialized and relaxed into a stable skyrmion state on ferromagnetic nanostrips (1000 nm × 100 nm × 1 nm) having inter-layer RKKY-type exchange coupling between them. The parameters utilized to stabilize the skyrmions in the ground state are as follows:[46] saturation magnetization for two layers of Co, $Ms_1 = 0.58$ MA/m and $Ms_2 = 0.58$ MA/m; exchange stiffness constant A = 15 pJ/m; iDMI at the Co/Pt interface D = 3.0 mJ/m²; uniaxial anisotropy in the z-direction $K_z = 0.8$ MJ/m³; Gilbert damping parameter $\alpha = 0.01$; and the RKKY interaction between the two FM layers $\sigma_{\mathrm{RKKY}} = -0.2$ mJ/m². By modulating the iDMI and the interlayer coupling, SAF skyrmions are stabilized. The skyrmions are initialized and relaxed into a stable skyrmion state at $x = 100$ nm in the nanostrip at the top layer and the bottom layer, as shown in Figure 1(b). The two ferromagnetic sublayers of the SAF are given the same iDMI values in the micromagnetic model, indicating symmetric HM/FM interfaces. The zoomed version of the two skyrmions stabilized at the top, and the bottom ferromagnetic layers are given by Figures 1(c) & (d), respectively. All micromagnetic simulations are conducted at 0 K, with thermal fluctuations disregarded. This method enables the isolation of the inherent effects of SOT, RKKY coupling, and iDMI on the stability and dynamics of SAF skyrmions without the need for extra stochastic contributions. As a result, the simulations provide the highest performance limit of the suggested device design and depict the idealised behavior of the system.

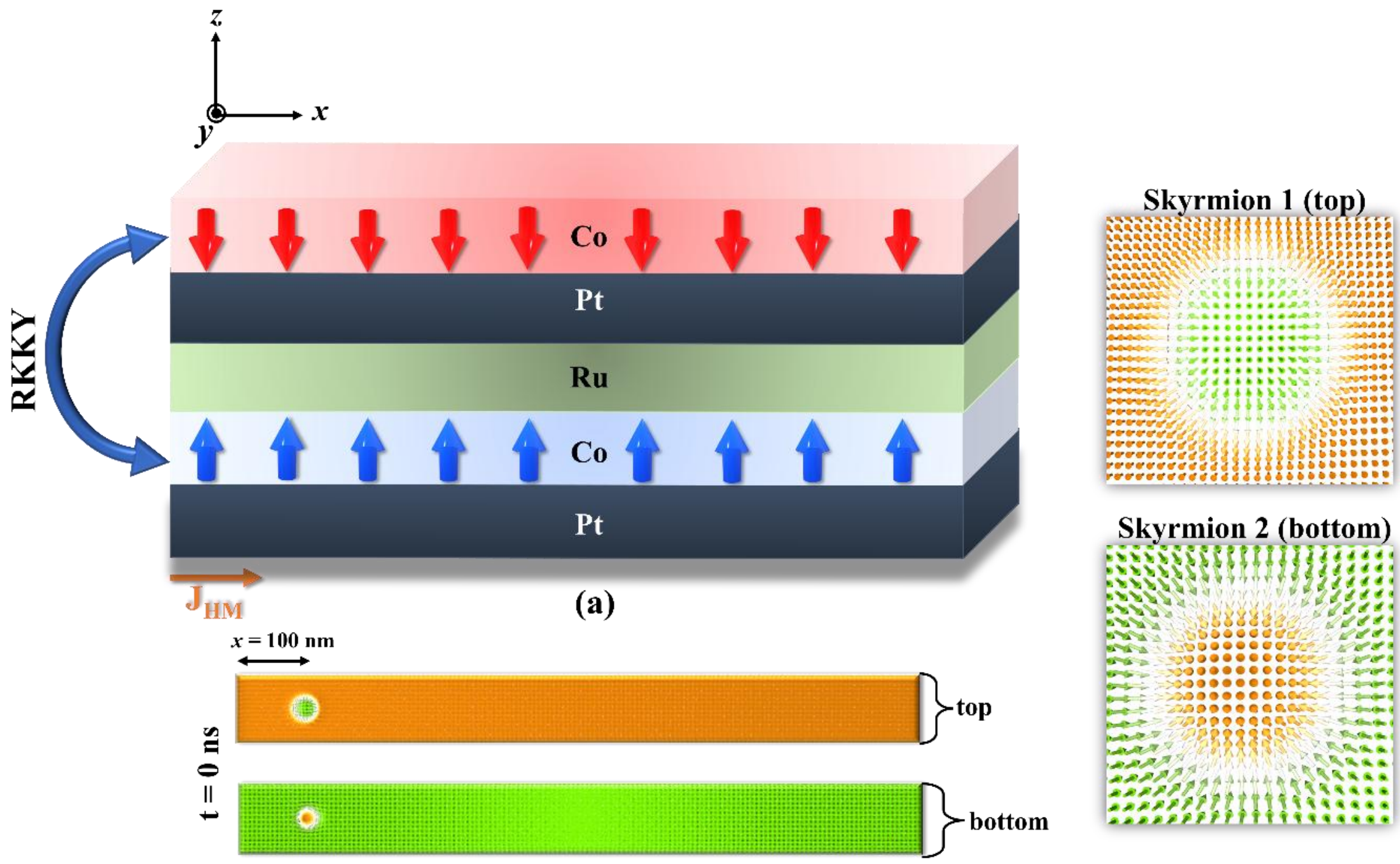

**Fig. 1 (a) A schematic of the multilayer structure consisting of Pt/Co/Ru/Pt/Co under consideration; the ferromagnetic Co layers are coupled antiferromagnetically by RKKY exchange via the Ru spacer. Applying a charge current $J_{HM}$ in the +x-direction in the Pt layers results in spin currents perpendicular to the interfaces via the spin Hall effect, with spin polarization p ∥ $-\hat{y}$ (right-handed coordinate system). Both ferromagnetic layers experience damping-like spin–orbit torques as a result of symmetric spin injection from the nearby Pt layers. (b) Ground state spin configuration and initial position of the two magnetic skyrmions at $x = 100\ nm$ on two ferromagnetic layers at $t = 0$. (c-d) Magnified images of the skyrmion spin textures in the top and bottom ferromagnetic nanostrips, respectively, which emphasize the mirrored Néel-type skyrmions arising from the antiferromagnetic interaction. The coordinate axes and the color bar showing the z-component of magnetization are also shown. The dimensions are not to scale.**

Simulations are performed to assess the static magnetic stability and structural properties of SAF skyrmions by systematically varying the iDMI and the interlayer RKKY exchange coupling strength, while maintaining all other parameters constant. For each arrangement, the system is allowed to attain its equilibrium condition, and the skyrmion radius is calculated independently for the upper and lower ferromagnetic layers. Through these simulations, we can quantify the impact of DMI and coupling strength on skyrmion size asymmetry across layers and construct a parameter space map that identifies regions of stable skyrmion configurations. The skyrmions initialized in the upper and lower ferromagnetic layers may also be propelled along the nanostrip by sending a current parallel to the plane of the multilayer structure. This current produces an SOT that propels the skyrmions. In the simulations, the SOT parameters are characterized by a spin Hall angle $\theta_{\mathrm{SH}} = 0.15$ and a spin polarization factor of $0.15$. The symmetric current injection from the neighboring Pt layers in the multilayer stack is represented by the uniform application of the SOT to both ferromagnetic layers.

The positive spin Hall angle results in the spin polarization vector $\boldsymbol{p}$ of the pure spin current being aligned along the negative $y$-axis, in accordance with the orientation of spins introduced into the ferromagnet by the spin Hall effect. The magnitude of the pure spin current injected into the ferromagnetic (FM) layer through the spin Hall effect is expressed as $|J_{spin}| = |\theta_{SH}|.|J_{HM}|$, where $J_{HM}$ represents the charge current density in the heavy metal layer. The pure spin current traverses the positive z-axis, producing a damping-like SOT that drives skyrmion motion.[47] The dynamic response of SAF skyrmions is characterized by calculating the average velocity of skyrmions in the top and bottom layers as a function of applied current density within a specified range, for two distinct iDMI values. The equilibrium radius of the skyrmions is computed at each current density to evaluate any deformation resulting from current-induced dynamics. The data analyses are performed using a home-built code that monitors the profile of the skyrmion during the simulation. The SkHA has been determined at two DMI values to measure any transverse motion in the presence of SOT. Appendix A provides analytical expressions for the scaling of skyrmion radius with iDMI, current-driven velocity, and cancellation of gyrotropic forces in SAFs.

## Results and discussion

### A. Influence of iDMI and RKKY coupling strength: Static characteristics of SAF skyrmions

We first systematically investigate the static stability of SAF skyrmions within Pt/Co/Ru/Pt/Co multilayer by varying two critical parameters: iDMI and the interlayer exchange coupling strength ($\sigma_{RKKY}$). The stability of skyrmion textures in synthetic bilayers is essentially dependent on the interaction between these two parameters, which together dictate the chirality, size, and endurance of the resultant magnetic configurations.

To elucidate the role of iDMI in determining the static characteristics of skyrmions in SAF nanostrips, we conduct micromagnetic simulations by systematically varying the iDMI strength from 1 mJ/m² to 4 mJ/m² in 0.1 mJ/m² steps. To determine the stability window of SAF skyrmions, we establish a phase diagram by changing the iDMI strength while maintaining the RKKY coupling constant at $\sigma_{RKKY}$ = -0.2 mJ/m², as shown in Fig. 2(a). The DMI is not enough to balance the exchange and anisotropy energy for low iDMI values (D < 2.4 mJ/m²), and therefore, no skyrmion state is stabilized. The stable formation of skyrmion pairs in the upper and lower ferromagnetic layers is shown by our simulations under conditions of positive and sufficiently high iDMI, ranging from 2.4 mJ/m² to 3.7 mJ/m² with a constant coupling strength of $\sigma_{RKKY}$ = –0.2 mJ/m². Beyond this regime, the skyrmion structures undergo increased deformation at larger iDMI values, which ultimately results in deformation or instability due to boundary-induced effects and excessive chiral twisting. This creates a clearly defined stability window for SAF skyrmions in the multilayer geometry under consideration. All the simulations are performed in the absence of an external magnetic field to isolate the intrinsic effects arising solely from iDMI. All static states are also derived using full-time integration of the LLG equation, instead of employing `relax()` or `minimize()` functions, to guarantee precise equilibrium configurations. A large damping coefficient (α = 0.9) is employed for 20 ns without applied current to expedite convergence. The damping is subsequently reduced to 0.01 for dynamic simulations. Under these circumstances, two Néel-type skyrmions with opposite topological charges are stabilized in the SAF structure, resulting in a net-zero topological charge and magnetization. We observe a substantial increase in skyrmion radius with increasing iDMI: at D = 2.4 mJ/m², the radii in the upper and lower layers are 2.76 nm and 3.19 nm, respectively, whereas at D = 3.7 mJ/m², they increased to 22.17 nm and 22.21 nm, respectively. This is demonstrated in Figure 2(b) for skyrmion 1 (located on the upper ferromagnetic layer) and skyrmion 2 (located on the bottom ferromagnetic layer). Although the iDMI is similar in both layers, small variations in equilibrium skyrmion radii occur due to layer-specific effective fields, including dipolar and boundary effects. The significant variation at $D$ = 2.4 mJ/m² is ascribed to finite-size and magnetostatic boundary effects near the lower stability threshold, rather than to iDMI variations. Throughout the entire iDMI range, the skyrmion profiles remain stable, and the antiferromagnetic alignment between ferromagnetic layers is preserved, confirming that iDMI serves as an important tuning parameter for skyrmion size in SAFs, with important implications for device scalability and energy-efficient spintronic manipulation. The skyrmion size in both layers increases with increasing iDMI strength, and the radii converge closely with increasing iDMI strength, indicating improved symmetry of the skyrmion pair in the compensated SAF configuration. To generalize the stability criterion, we construct a phase diagram by systematically varying the iDMI strength and interlayer RKKY coupling (Fig. 2(c)). Stable skyrmions are not achieved for negative iDMI if the original chirality is maintained as in the positive-D scenario. Reversing the chirality to align with the favored rotational direction of negative iDMI reinstates stable SAF skyrmion pairs. This verifies that the observed sign dependency originates from chirality compatibility rather than the intrinsic DMI sign itself. Chirality in single-layer ferromagnets is primarily determined by the iDMI sign, but stability is dependent on its magnitude compared to anisotropy and exchange energies.[3] Conversely, SAF systems introduce an additional constraint via antiferromagnetic interlayer interaction, necessitating antiparallel alignment among layers. Stability is only attained when the iDMI-favored chirality in each layer is consistent with this interlayer alignment, as previously discussed in AFM/SAF studies.[20,23] Otherwise, frustrated spin textures result in instability.

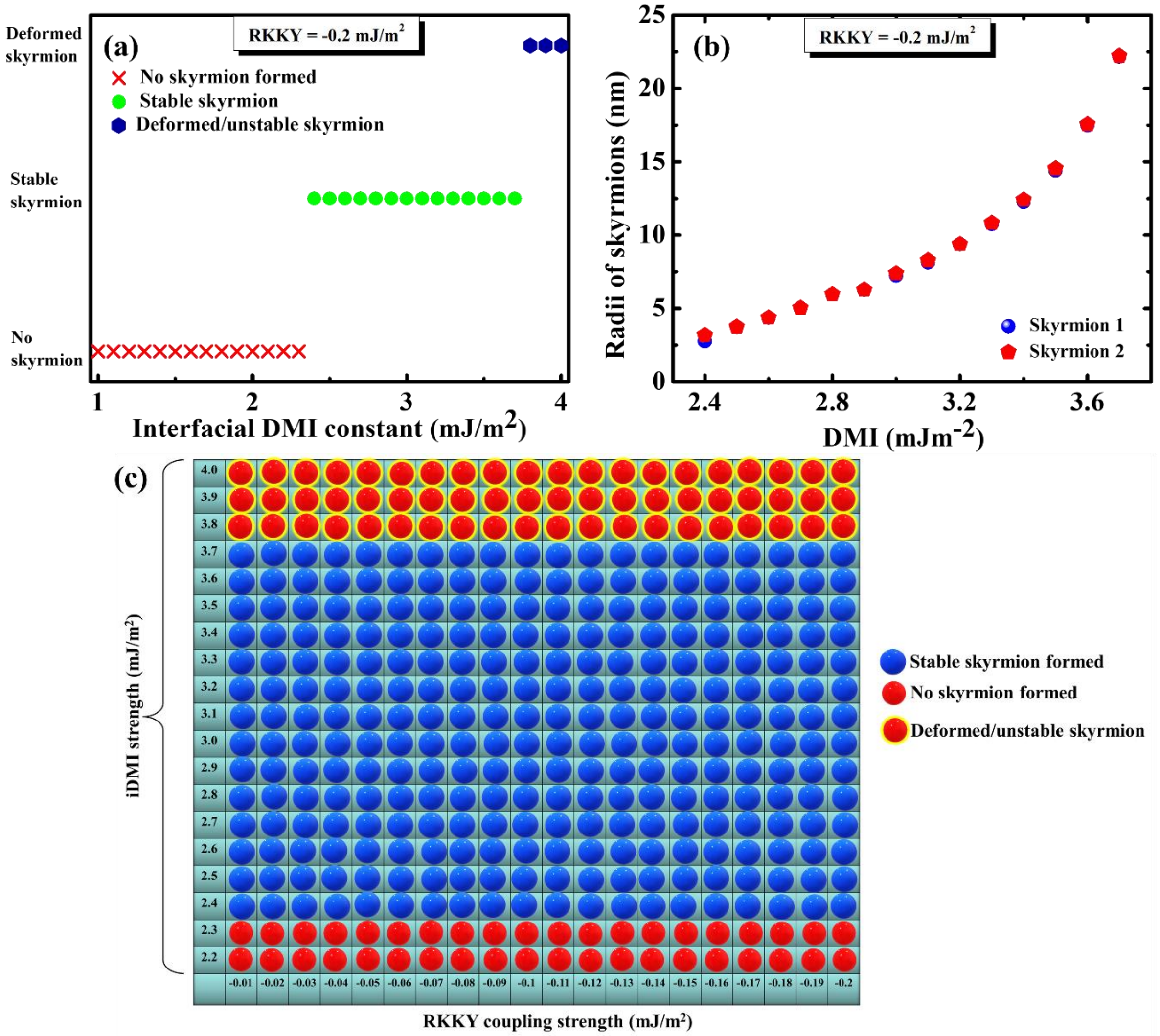

**Fig 2. (a) SAF skyrmion stability diagram with a given antiferromagnetic interlayer coupling $\sigma_{RKKY}$ = -0.2 mJ/m² as a function of iDMI (b) Variation in skyrmion radius on the top ferromagnetic layer (skyrmion 1; blue) and bottom ferromagnetic layer (skyrmion 2; red) as a function of iDMI strength within the stable region, with the RKKY coupling fixed at –0.2 mJ/m² (c) Phase diagram in 2D showing the stability of SAF skyrmions as a function of iDMI and antiferromagnetic RKKY coupling strength. Blue symbols represent stable, coupled SAF skyrmion pairs; the lack of skyrmion formation is shown by red symbols, and unstable or distorted skyrmion configurations are indicated by red symbols underlined in yellow. The diagram depicts the parameter window necessary for effective stabilization of compensated skyrmion pairs.**

We subsequently vary the RKKY coupling from 0 to −0.2 mJ/m² (step −0.01 mJ/m²) for D = 3.0 and 3.5 mJ/m², within the stability window of Fig. 2(c), to assess its effect on skyrmion morphology. In both cases, only minor changes in radius are observed, and stable compensated skyrmion pairs persist across the entire range. For D = 3.0 mJ/m², the radii are ~7.22 nm (top) and ~7.34 nm (bottom), giving a mean of ~7.28 nm. For D = 3.5 mJ/m², they are ~14.42 nm and ~14.64 nm, with a mean of ~14.53 nm. As is also evident in Appendix A, increasing iDMI strength causes the skyrmion radius to expand. Interlayer RKKY coupling significantly influences the collective dynamics of SAF skyrmions by affecting the effective dynamical tensors and reducing gyrotropic forces.[47] However, for static radius estimates, the dominant effect is iDMI scaling, with RKKY providing only minor corrections (discussed in Supplementary Information). These findings show that the RKKY interaction mainly enforces antiparallel alignment and topological compensation without a substantial change in the intrinsic skyrmion size in the weak-to-moderate antiferromagnetic domain. The skyrmion radius is therefore primarily dictated by the balance of intralayer exchange, iDMI, perpendicular magnetic anisotropy, and magnetostatic energy, with interlayer exchange playing a supporting role to determine static morphology.

In contrast, under ferromagnetic coupling ($\sigma_{RKKY} > 0$), a distinct change in skyrmion configuration and radii is observed. Specifically, for $\sigma_{RKKY} = +0.2$ mJ/m$^2$ and D = 3.5 mJ/m$^2$, the radius of the skyrmion in the bottom layer (skyrmion 2) is reduced to 5.29 nm, compared to 14.54 nm under the same conditions in antiferromagnetic coupling. The skyrmions are initially stabilized in both the top and bottom ferromagnetic layers at a low coupling value of +0.01 mJ/m$^2$ when the RKKY exchange strengths vary from +0.01 mJ/m$^2$ to +0.2 mJ/m$^2$, incremented in steps of +0.05 mJ/m$^2$. Nevertheless, skyrmion generation is observed exclusively in the bottom ferromagnetic layer from a coupling strength of +0.05 mJ/m$^2$ onwards. As $\sigma_{RKKY}$ becomes positive, this behavior suggests that the system transitions from a compensated SAF regime to a ferromagnetically coupled regime. In this regime, the lack of a coupled skyrmion in the upper layer and the continued presence of an isolated skyrmion in the lower layer indicate a loss of mirror symmetry. The transition is further validated by the positive binding energy of the skyrmion pair ($E_{bind} = E_{coupled} - E_{uncoupled} > 0$), indicating that the bilayer structure becomes energetically unfavorable, and the system no longer sustains a bound SAF skyrmion state. The layer-asymmetric stabilization of skyrmions in coupled multilayer systems has been previously observed, wherein the interaction of interlayer exchange, dipolar fields, and iDMI results in the selective stabilization or collapse of skyrmions in distinct layers, rather than uniform textures across all layers.[24]

These findings collectively define the phase space for skyrmion stability in SAF multilayers. The modification of skyrmion radius and layer occupancy via RKKY strength and DMI facilitates the design of device-specific SAF skyrmions, suitable for robust topological logic, low Hall deflection racetrack motion, or synchronized phase oscillators in layered structures. Therefore, the static findings presented here serve as the basis for the dynamic behaviors covered in subsequent parts. Understanding the interrelation between DMI and RKKY coupling is crucial for customizing SAF skyrmions to achieve specific dimensions, stability features, and motion characteristics in advanced spintronic devices.

### B. Dynamic behavior of SAF skyrmions in response to SOT

In addition to static stabilization, the dynamic behavior of SAF skyrmions under SOT provides a profound understanding of their transport properties and prospective use in fast spintronic applications. This study carefully investigates the behavior of SAF skyrmions in response to varying spin current densities, focusing on their velocity, deformation, topological characteristics, and the feasibility of sustained motion in planar nanostrip configurations.

### *1. Velocity of SAF skyrmions as a function of current density*

To investigate the dynamic response of SAF skyrmions under SOT, we systematically vary the applied current density and measure the average velocity of the skyrmion in both the upper and lower ferromagnetic layers at two distinct DMI values, including PBC. The velocity is determined by monitoring the position of the center of the skyrmion core over time via a home-built post-processing code. In accordance with the hypothesis that the SOT strength scales linearly with the charge current in the heavy metal layer, the velocities show a monotonic rise with current density, irrespective of the DMI values that are investigated, and reach as high as $v = 380$ m/s when iDMI = 3.5 mJ/m² and $J_{HM} = 1 \times 10^{11} A/m^2$. Figure 3(a-b) illustrates the velocity of the two skyrmions with an iDMI value of 3.0 mJ/m², while Figure 3(c-d) depicts their velocity at an iDMI value of 3.5 mJ/m².

At reduced current densities, the instantaneous motion of skyrmions is sluggish and displays temporal velocity variations, indicating momentary inertial effects and insufficient momentum transfer, even in an ideal, disorder-free configuration. As the current density increases ($J$ ~$10^{10}$ A/m$^2$), these variations gradually decrease, resulting in more uniform motion, which signifies improved dynamic stability. At $J$ ~ $10^{11}$ A/m$^2$, $10^{12}$ A/m$^2$, the skyrmion motion becomes more streamlined, with greater and more uniform velocity, indicating steady-state translational transport. The skyrmion velocity rises consistently with current density; however, a little divergence from ideal linear scaling is seen at higher currents, especially with reduced DMI strengths, indicating the emergence of internal skyrmion deformation and increased dissipation. We determine a critical current density $J_c$ = 5×$10^7$ A/m², which is the minimum applied current necessary to maintain translational motion with a non-zero time-averaged velocity. For $J < J_c$, the skyrmions cease to move along the considered racetrack. However, the symmetric response of the skyrmions in the two ferromagnetic layers demonstrates strong antiferromagnetic coupling during propagation. Thus, the SAF configuration facilitates swift motion along nanotracks while mitigating edge-induced instabilities often seen in single-layer ferromagnetic systems.

Submitted on June 25, 2026

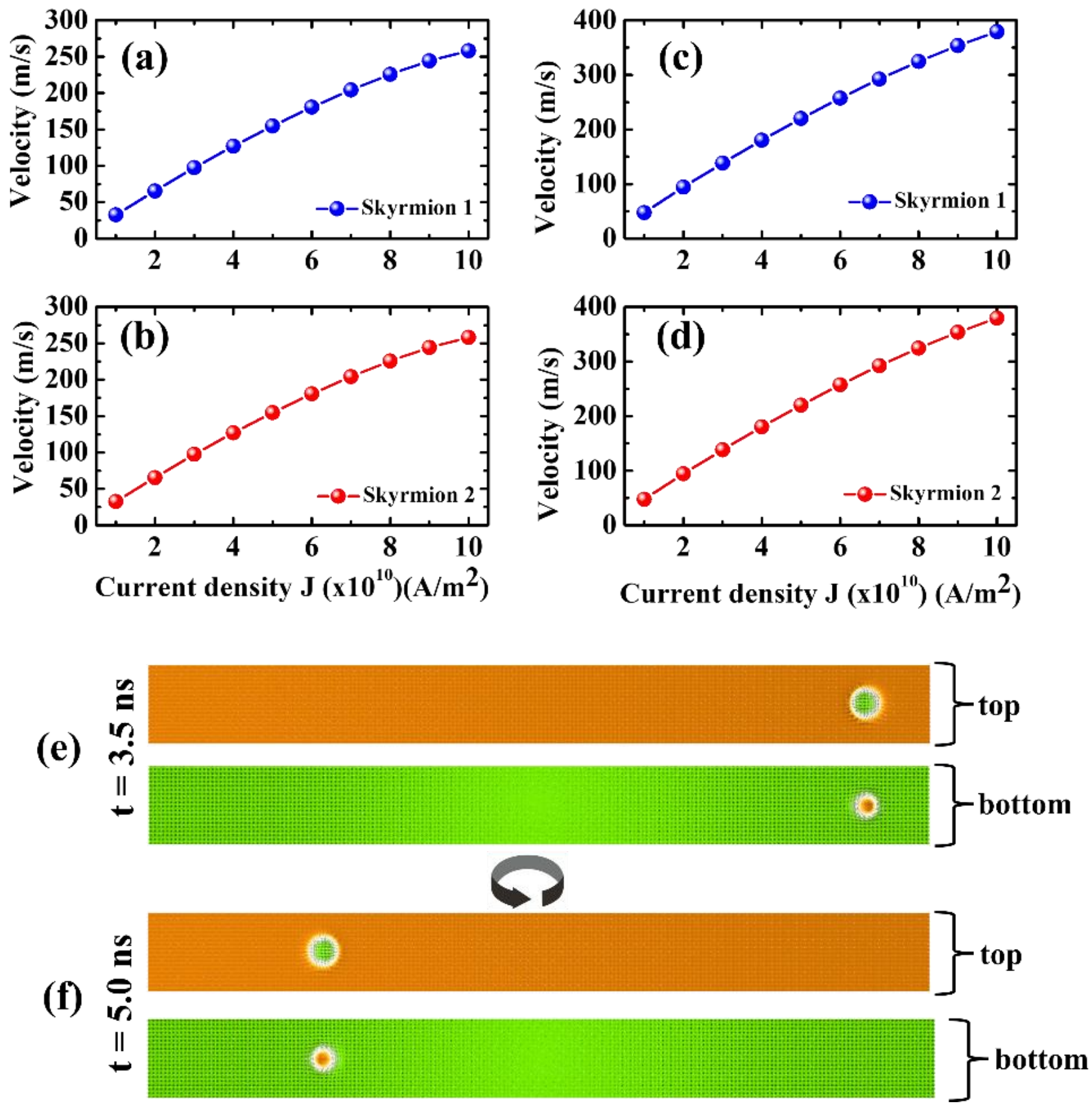

**Fig 3. Average skyrmion velocity in the top (blue circles) and bottom (red circles) ferromagnetic layers as a function of applied current density for iDMI strengths: (a-b) D = 3.0 mJ/m², (c-d) D = 3.5 mJ/m². In both cases, velocity increases monotonically with applied current density. (e-f) Snapshots of the motion of the magnetic skyrmions at two different times, showing the repetitive translational motion of the SAF skyrmions along the nanotrack.**

## *2. Recurrent translational dynamics of skyrmion pairs under periodic boundary conditions*

We explore the translational motion of the skyrmion pairs in the SAF nanotrack under both PBC along the *x*-direction (infinite nanotrack) as well as in confined nanotracks to investigate their dynamics and potential applications in time-domain devices such as an oscillator or a clock. When modeled with PBC, this transport represents a closed-loop configuration where consecutive traversals establish a distinct temporal period determined by the nanotrack length and skyrmion velocity. This approach offers a practical theoretical foundation for assessing long-range, steady-state skyrmion transport without the need for geometric curvature or additional confinement potentials. To capture the intrinsic transport characteristics of SAF skyrmions, closed-loop or effectively infinite racetrack geometries are modeled using PBCs, which eliminates edge-induced effects and serves as an idealized framework.[30,48] This approach eliminates undesirable reflections and avoids additional confinement or geometry-induced distortions. Finite nanotrack simulations without PBC are also performed for completeness. Although SAF skyrmions retain similar longitudinal motion, edge interactions induce deformation, slight transverse drift, and interlayer symmetry breaking. At moderate RKKY coupling (−0.2 mJ/m²), this leads to rebound-like dynamics, whereas stronger coupling (−0.5 mJ/m²) enforces rigid interlayer locking and suppresses recurrence near the edges. These effects originate from edge-induced asymmetry, which disrupts the cancellation of the gyrovector. Thus, while boundary effects lead to deviations in finite systems, PBCs are indispensable for isolating intrinsic, symmetry-preserving dynamics. The observed recurrence is not a consequence of PBCs, but rather a manifestation of true topology-compensated spin transport inherent to closed-loop geometries.

In practical scenarios, similar recurrent transport may be accomplished in annular racetracks, where SAF skyrmions continuously revolve over a closed-loop path. Each complete traversal of the loop constitutes a timing event, with the corresponding frequency tunable via material parameters and the applied current density. The suggested clock design concept is compatible with experimentally demonstrated Pt/Co/Ru/Pt/Co SAF multilayers that support sustained skyrmions due to the combined effects of iDMI, PMA, and antiferromagnetic RKKY coupling. A realistic implementation can employ an annular racetrack patterned from the SAF multilayer, with a circumferential current introduced via a surrounding heavy-metal ring to generate a tangential SOT that propels skyrmion motion. Local magnetic tunnel junctions (MTJs) positioned along a racetrack can detect the periodic motion of the SAF skyrmion through variations in magnetoresistance, thereby transforming repetitive translational motion into an electrical clock signal.[30,49,50] Since the necessary material parameters are within the experimentally reported ranges for SAF skyrmion systems, the proposed device concept should be compatible with current spintronic fabrication methods and characterization techniques. Figure 3(e-f) shows the motion of the pair of skyrmions at two subsequent times, t = 3.5 ns and t = 5 ns, respectively. Under PBC, the skyrmions return to the computational domain after one longitudinal traversal, recreating their initial spatial arrangement. This repetitive motion, maintained over several nanoseconds, is robust to annihilation, thereby establishing a dynamic equilibrium along the nanostrip that is characterized by continuous, periodic translational motion. We also perform micromagnetic simulations with PBC, having nanostrip lengths of 2000 nm and 4000 nm, run for 20 ns and 40 ns, respectively. Despite these changes in longitudinal dimensions, the results remain unchanged. In contrast to ferromagnetic skyrmions, which undergo a lateral drift caused by the skyrmion Hall effect and are vulnerable to edge annihilation, SAF skyrmions with opposite topological charges encounter no net Magnus force. This results in a pure longitudinal motion in the direction of the current, enabling stable, repetitive trajectories without requiring additional geometric confinement or external magnetic fields. In contrast to skyrmion oscillators that depend on breathing or gyration, the oscillation mechanism described here is exclusively translational, resulting from the recurrent longitudinal motion of SAF skyrmions with compensated topological charge. The controlled and confined oscillatory dynamics are inherently distinct from the unidirectional drift often observed in single-layer ferromagnetic systems and suggest an effective platform for skyrmion-based timing devices or nano-oscillators. The emergence of this periodic translational motion is governed by a critical current density $J_c$, which defines the activation threshold of the system. For $J < J_c$, sustained loop traversal is absent, whereas for $J \geq J_c$, stable recurrent motion emerges. In this regime, each full traversal of the annular track corresponds to a distinct temporal cycle. Such controlled periodic dynamics are well-suited for spintronic timing elements, where each oscillation cycle effectively represents a clock pulse, enabling current-tunable time-keeping functionality.

In tracks with open boundaries, we implement a high-damping area ($\alpha = 0.95$) at one end ($x \geq 950\ nm$) to act as a soft barrier that gradually decelerates the skyrmions without causing their annihilation. Upon entering a high-damping region ($\alpha = 0.95$), the skyrmion loses kinetic energy, resulting in deceleration and a modification of its trajectory. It experiences a brief damped oscillation at the boundary with progressively decreasing amplitude, eventually stabilizing rather than sustaining continuous motion.

### 3. *Modulation of skyrmion radii with current density*

Apart from velocity, the change in skyrmion size is also investigated with the increase in current density. The skyrmion radii are measured along the nanostrip in both ferromagnetic layers for two distinct iDMI values. The skyrmion radii gradually decrease as the current density increases, indicating greater compression of the spin structure under enhanced SOT. The effect is more significant in the bottom ferromagnetic layer due to direct spin injection, whereas the upper ferromagnetic layer maintains a marginally greater radius, which corresponds with its indirect stimulation, as shown in Figure 4.

The reduction in skyrmion radius with increasing current density contrasts with intrinsic antiferromagnetic skyrmions, which typically exhibit current-induced expansion.[51,52] The difference stems from distinct underlying mechanisms: intrinsic AFMs experience size expansion due to internal excitations and sublattice canting, whereas in SAFs, damping-like SOT alters the domain-wall configuration, leading to a reduced effective radius.

Again, the influence of iDMI strength on radius evolution is clear. While deformation becomes more pronounced at lower iDMI (D = 3.0 mJ/m²), the skyrmion retains more stiffness and is less compressible at higher iDMI (D = 3.5 mJ/m²). For D = 3.0 mJ/m², the average radius in both the top and bottom ferromagnetic layers is approximately 7.4 nm and 7.3 nm, respectively. Beyond a current density of $7 \times 10^{10}$ A/m², the radii of the moving skyrmions decrease sharply, indicating a nonlinear deformation response. Similarly, for D = 3.5 mJ/m², the average radii are around 14.8 nm and 14.7 nm for top and bottom layers, with a rapid decline observed beyond the same current density. Thus, high-speed transport favors smaller, more compact skyrmion designs due to a more dynamic energy landscape. These observations highlight the sensitivity of skyrmion morphology to SOT-driven dynamics and underline the role of iDMI strength in modulating skyrmion stability under high current densities. For devices that necessitate variable skyrmion size, this variation provides a tuning parameter. The size reduction is symmetrically maintained in both FM layers, further emphasizing the versatility of the SAF design during dynamic performance.

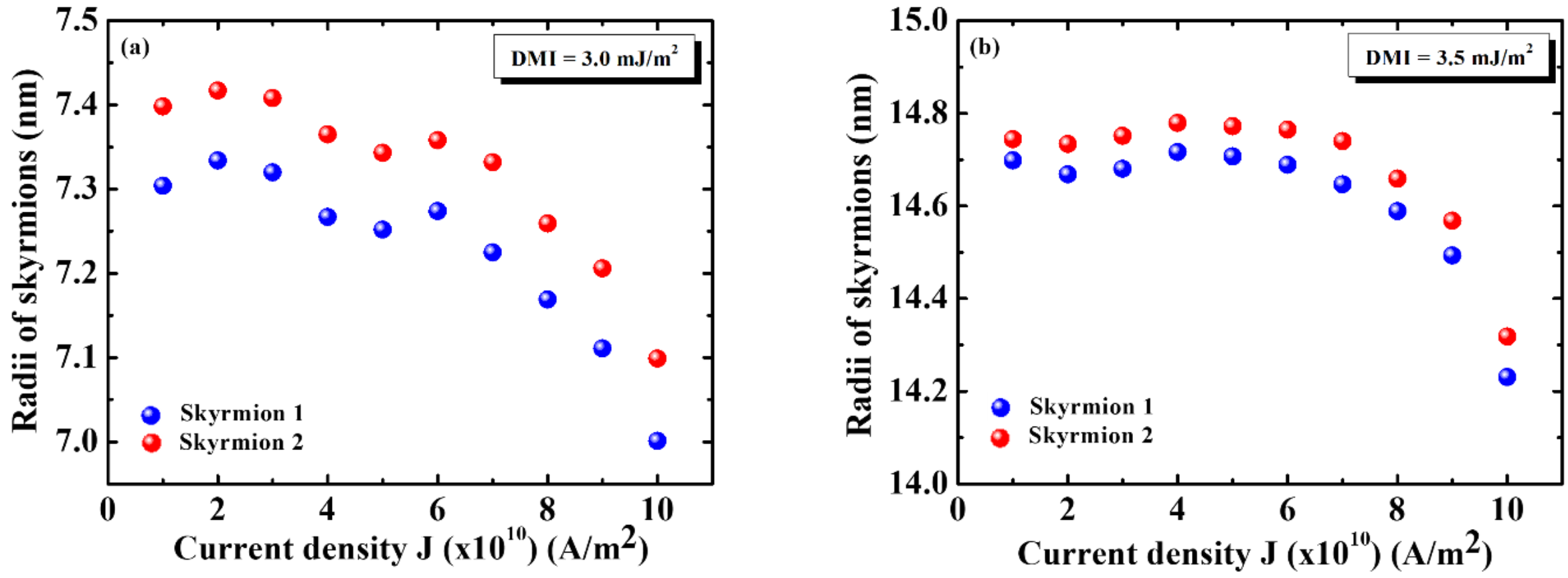

**Fig 4. Variation in skyrmion radii for the top ferromagnetic layer (skyrmion 1) and bottom ferromagnetic layer (skyrmion 2) as a function of current density, with RKKY coupling fixed at –0.2 mJ/m². Panel (a) corresponds to a fixed iDMI value of 3.0 mJ/m², while panel (b) shows results for iDMI = 3.5 mJ/m².**

### *4. Suppression of skyrmion Hall angle and cancellation of topological charge*

SAFs have emerged as a promising platform for skyrmion-based spintronic devices, offering a viable route to overcome the limitations imposed by the SkHE in planar ferromagnet counterparts. In conventional ferromagnets, skyrmions exhibit transverse motion under the influence of SOT due to their non-zero topological charge. In contrast, SAFs consist of ferromagnetic layers coupled via RKKY interaction, enabling the formation of skyrmions with compensated topological charge. This compensation effectively suppresses the SkHA and allows for linear, guided motion. The cancellation of net topological charge suppresses the gyrotropic (Magnus) force and supports streamlined, fast, and low-power transport, making SAFs well-suited for racetrack memory and logic applications.

In this work, we analyze the trajectory and the transverse deflection of skyrmions in a SAF nanostrip by computing the SkHA across a variety of current densities. The analyses are performed for two distinct iDMI values. Our study of SkHA demonstrates that the SAF configuration significantly eliminates transverse deflection, even under high current densities. The SkHA remains approximately zero across the studied parameter space. For D = 3.0 mJ/m², the minimum deflection angle observed at a current density of $1 \times 10^{10}$ A/m² was 0.033°, while the maximum angle reached 0.203° at $1 \times 10^{11}$ A/m² for skyrmion 1. In the case of D = 3.5 mJ/m², the deflection ranged from 0.023° to 0.205° over the same current range. Increasing iDMI enhances domain-wall chirality and expands the skyrmion radius, thereby making the spin texture more resilient to current-induced deformation. In SAFs, opposite topological charges of the coupled skyrmions effectively cancel the net gyrovector, maintaining a negligible skyrmion Hall angle. The small increase in transverse deflection seen at higher iDMI is not because of a substantial change in the effective $\frac{G}{D}$ ratio but instead arises from subtle asymmetries in dissipative contributions and layer-specific dynamic responses. Despite the internal skyrmion structure evolving with increasing iDMI strength, the compensated topology preserves predominantly longitudinal motion with negligible Hall deflection. Figure 5 displays the plots of the SkHA measured at different current densities at two iDMI values. These results indicate minimal transverse displacement, confirming effective suppression of SkHE. Additionally, we calculate the topological charge density $q\ (x, y)$, and found that the top and bottom layers exhibit nearly equal and opposite values, further validating the cancellation mechanism. [The topological charge density can be defined as: $q\ (x, y) = \frac{1}{4\pi}\boldsymbol{m}(x, y) \cdot \left(\frac{\partial \boldsymbol{m}}{\partial x} \times \frac{\partial \boldsymbol{m}}{\partial y}\right)$, where $\boldsymbol{m}(x, y)$ is the unit magnetization vector.] For D = 3.0 mJ/m², the topological charge density of skyrmion 1 is $\langle q\ (x, y)\rangle$ = -4.63×$10^{12}$/ m² while it is $\langle q\ (x, y)\rangle$ = +4.64 ×$10^{12}$/m² for skyrmion 2. While, for D = 3.5 mJ/m², the values are $\langle q\ (x, y)\rangle$ = -4.57×$10^{12}$/m² and $\langle q\ (x, y)\rangle$ = +4.58 ×$10^{12}$/m² respectively. The calculated values represent the spatially averaged topological charge density, obtained by averaging the local topological charge density over the entire ferromagnetic layer, under both static and current-driven conditions. The approximately equal and opposite averaged densities correspond to integrated topological charges Q $\approx \pm 1$, confirming the compensated topology of the SAF skyrmion pair. The persistence of this compensation in both static and current-driven scenarios further validate the suppression mechanism of the skyrmion Hall effect. As the iDMI intensity increases, the skyrmion radius enlarges owing to increased chiral stability of the domain wall. Consequently, the spin rotation encompasses a broader spatial area, resulting in an expanded radial distribution of the local topological charge density $q\ (x, y)$. In contrast to smaller D values, which provide a more compact skyrmion with spin twisting confined to a more compact area, greater D results in a broader spatial profile of $q\ (x, y)$ concentrated around the enlarged skyrmion border. Despite variations in local distribution with D, the overall integrated topological charge Q of each layer remains quantized and invariant. At current density values considered in this work, the skyrmion exhibits translational motion without much distortion of its internal spin configuration. Thus, the spatial distribution of $q\ (x, y)$ is maintained during propagation, and the overall topological charge stays invariant until a topological transition, such as annihilation or merger, occurs. The preservation of topological charge during motion demonstrates the dynamical resilience of the SAF skyrmion, hence facilitating reliable functionality for logic, memory, and clock-driven device applications. The current work focuses on the deterministic dynamics of SAF skyrmions at 0 K in perfect defect-free structures. Thermal fluctuations may affect the long-term transport behavior, depinning thresholds, deformation, and stability of the skyrmions. However, because of its topological origin, the suppression of the skyrmion Hall effect resulting from topological charge compensation is anticipated to be resilient despite modest thermal perturbations. However, the inclusion of defects and pinning centres (e.g., grain disorder) will be crucial for realistic device modelling and optimization.

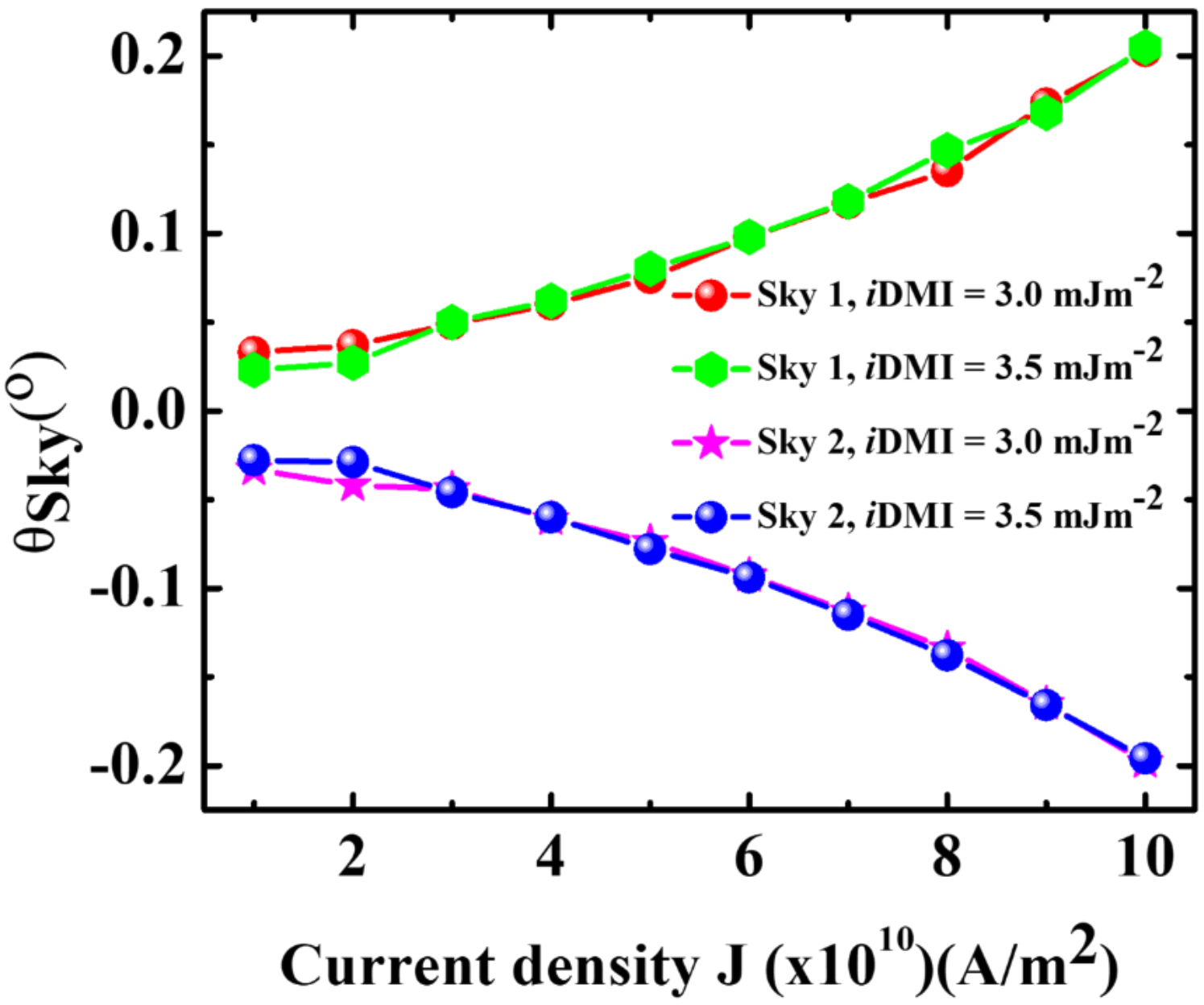

**Fig 5. Plots of Skyrmion Hall Angle as a function of current density strength for two different iDMI values.**

### 5. *Recurrent SAF skyrmion dynamics: Device Implications*

This study illustrates that the recurrent dynamics of SAF skyrmions allow a device architecture where translational skyrmion transport is immediately translated into periodic electrical signals via localized magnetoresistive detection. Due to the cancellation of gyrotropic forces and the consequent suppression of the skyrmion Hall effect, SAF skyrmions demonstrate robust longitudinal propagation with low transverse displacement and deformation. These attributes are especially appealing for low-power spintronic interconnects, signal-distribution networks, and timing systems necessitating reliable transmission over long distances.

Employing PBC to model an annular racetrack, a circulating SAF skyrmion pair continuously passes over a specified detecting region, such as an MTJ, producing a series of periodic timing events. The operational frequency of the proposed SAF-skyrmion clock is defined by the ratio of the current-dependent skyrmion velocity to the effective propagation length of the closed loop, $f(J) = v(J)/L_{eff}$. The skyrmion velocity approximately increases linearly with current density in the steady-flow region, allowing for continuous tuning of the operating frequency by electrical control. The material properties investigated in this study indicate that the expected operation frequencies range from sub-GHz to GHz, with further scaling possible via device miniaturization and geometric optimization. Figure 6 depicts the proposed clock design and the physical interpretation of the PBCs used in the simulations. The periodic nanotrack used in the computational model resembles an annular racetrack geometry, whereby the SAF skyrmion perpetually revolves under SOT. Each transit over a predefined MTJ readout region results in a timing event, which converts recurring translational motion into a periodic electrical clock signal.

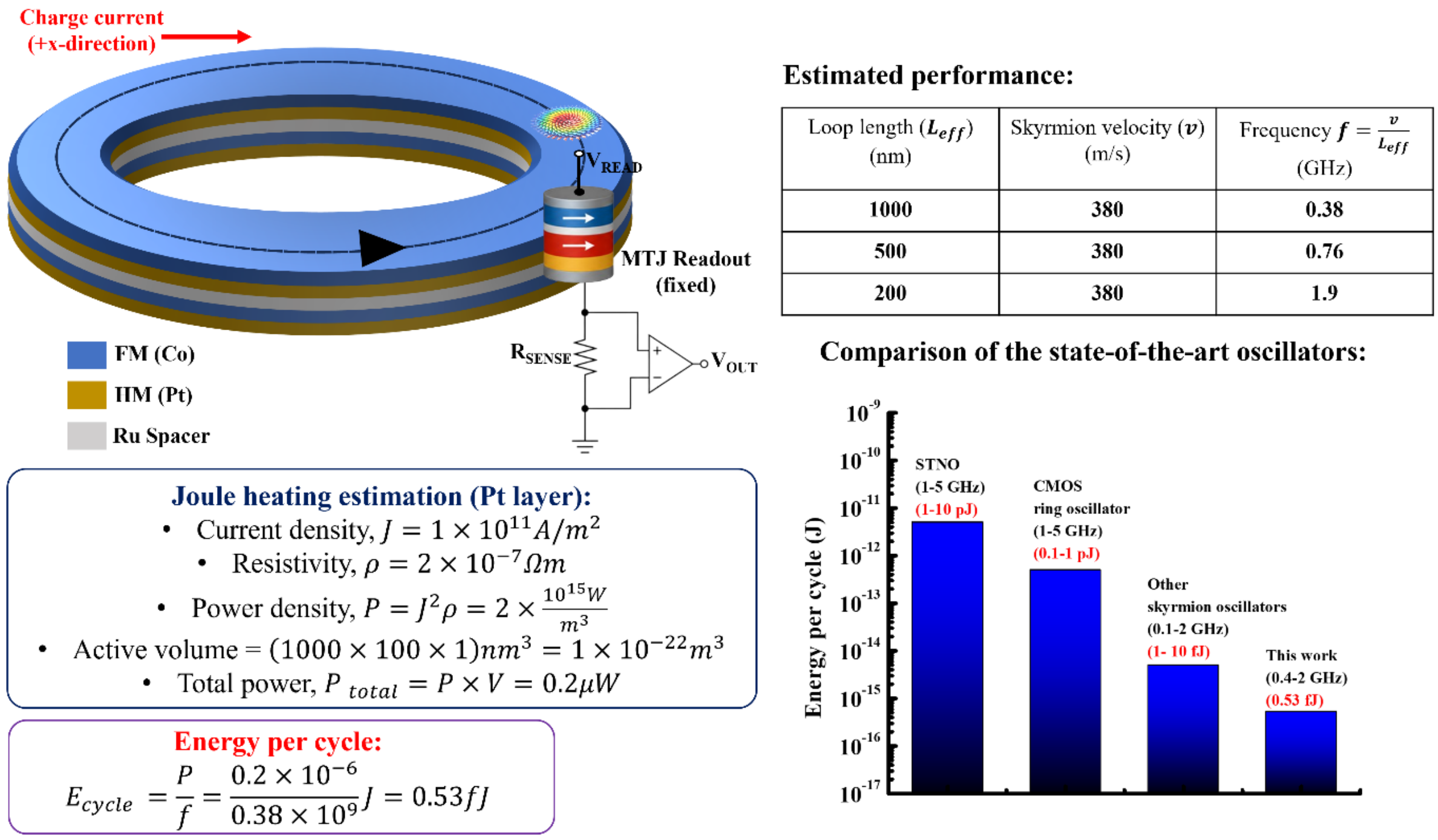

**Fig 6. Proposed SAF Skyrmion Clock: Device Concept and Estimated Performance.**

The initiation of recurrent transport is determined by a critical current density, $J_c$. For $J < J_c$, the applied SOT is inadequate to maintain long-range propagation, hence inhibiting periodic signal creation. When $J \geq J_c$, reliable recurrent transport emerges, generating periodic electrical pulses upon detection. This threshold-activated transition mimics tonic-spiking behavior in biological neurons,[53] whereby the firing frequency escalates with stimulus strength. Thus, the resultant frequency-agile pulse trains may be used for temporal encoding, spike-based information processing, oscillator-based computing, and neuromorphic architectures.

This working concept is substantially different from current spintronic oscillator technology. Spin-transfer torque nano-oscillators (STNOs) produce periodic signals by continuous magnetization precession, whereas vortex oscillators depend on gyrotropic vortex-core dynamics. Likewise, previous reported skyrmion oscillators mostly use breathing modes, gyration dynamics, or resonant internal excitations of magnetic skyrmions. Conversely, the periodic response described herein arises from the repetitive translational motion of a topologically compensated SAF skyrmion pair. As a result, instead of inherent magnetic eigenmodes, the oscillation period is controlled by the skyrmion traversal time around a closed-loop racetrack, offering a unique mechanism for signal generation and clocking.

Additionally, the analytical framework outlined in Appendix A offers on the scaling of skyrmion size with material parameters and the suppression of gyrotropic forces through topological charge compensation. The analytical estimates presented in Appendix B predict a sub-femtojoule energy consumption per clock cycle, underscoring the energy-efficient nature of the proposed architecture. The analytical and micromagnetic findings together provide a robust foundation for the design of frequency-tunable SAF-skyrmion-based timing elements, signal generators, synchronization units, and prospective low-power neuromorphic spintronic systems.

Table 1. Comparison of the proposed SAF-skyrmion clock to typical spintronic oscillator architectures.

| **Oscillator Architecture** | **Oscillation Mechanism** | **Field-Free Operation** | **Frequency Tunability** | **References** |
|---|---|---|---|---|
| STNO | Magnetization precession | Partial | High | [54] |
| Vortex oscillator | Vortex-core gyration | Partial | Moderate | [55] |
| Breathing skyrmion oscillator | Skyrmion radius oscillation | Device-dependent | Moderate | [56] |
| Gyrating skyrmion oscillator | Orbital/gyration motion of skyrmions | Device-dependent | Moderate | [30, 57] |
| **Proposed SAF-skyrmion clock** | **Recurrent translational motion** | **Yes** | **High** | **This work** |

# Conclusion

In summary, this study identifies SAF skyrmions as viable candidates for dynamic spintronic device applications, extending their conventional role as information carriers. Utilizing systematic micromagnetic simulations, we determined parameters necessary for stabilizing topologically compensated SAF skyrmions in Pt/Co/Ru/Pt/Co multilayers and proved their robust current-driven transport under SOT. The interplay of positive iDMI and antiferromagnetic RKKY coupling facilitates the generation of stable skyrmion pairs with opposite topological charges, leading to the suppression of the SkHE and ensuring highly reliable longitudinal propagation.

From a device standpoint, the most significant outcome of this study is the demonstration of recurrent translational motion as a mechanism for generating periodic signals. Utilizing PBC as an idealized mimic of an annular racetrack, we demonstrate that a circulating SAF skyrmion can continuously traverse a specified detection region, generating a series of timing events that can be transformed into electrical clock pulses through conventional magnetotransport readout methods. In contrast to traditional spintronic oscillators that use magnetization precession, vortex gyration, or skyrmion breathing modes, the proposed technique depends on the stable translational recurrence of a topologically compensated magnetic texture. The operational frequency is readily adjustable via current density and material characteristics, offering a straightforward method for electrically programmable nanoscale clock generation.

The demonstrated transport properties also suggest possible applications in oscillator-based computer systems, synchronization networks, and low-power spintronic interconnects. The threshold-activated initiation of recurrent motion and the production of repetitive pulse trains indicate potential for neuromorphic signal processing, whereby temporal encoding and spike-like dynamics are essential. The analytical framework in Appendix A demonstrates the relationship between skyrmion size and material properties, as well as the suppression of gyrotropic forces through topological charge compensation in SAF skyrmions, while the analytical estimates in Appendix B suggest sub-femtojoule energy consumption per clock cycle, highlighting the energy efficiency of the proposed device design. Together with the micromagnetic simulations, these findings provide a thorough theoretical and computational framework for understanding, designing, and optimizing frequency-tunable SAF-skyrmion-based timing elements, signal generators, and next-generation low-power spintronic technologies. Overall, this study presents a translational SAF-skyrmion clocking concept that integrates topological robustness, field-free operation, frequency tunability, and energy-efficient transport in a scalable planar configuration. The findings not only pave the way for skyrmion-based timing and synchronization devices but also illustrate how topology-compensated magnetic textures can be utilized to accomplish dynamic functionalities that surpass mere information storage, thereby laying the groundwork for future low-power spintronic and neuromorphic technologies. Further investigation that includes thermal fluctuations, material disorder, and practical readout methodologies will be crucial for evaluating device performance in realistic operating conditions and guiding experimental implementation.

# APPENDIX:

## A. Analytical Study of Skyrmion Radius and Collective Dynamics in a SAF[58-62]

### A1. Micromagnetic Energy Functional and Estimation of Variational Radius

We start with the conventional micromagnetic energy density for an ultrathin ferromagnetic film with iDMI, first articulated within the continuous theory of chiral magnetic solitons.

The overall energy estimates

$$E = d^2r[A(\nabla \boldsymbol{m})^2 + K(1-m_z^2) + D(m_z\nabla\cdot\boldsymbol{m} - \boldsymbol{m}\cdot\nabla m_z)] \quad (1)$$

where $A$ is the exchange stiffness constant, $K$ is the perpendicular magnetic anisotropy, $D$ is the iDMI constant, and $\boldsymbol{m}$ is the unit magnetization vector.

### A2. Variational Ansatz (Néel Skyrmion Profile)

Let us consider an axisymmetric skyrmion:

$$\boldsymbol{m}(r,\phi) = (sin\theta(r)cos\phi, sin\theta(r)sin\phi, cos\theta(r)) \quad (2)$$

with the profile of the domain wall

$$\theta(r) = 2tan^{-1}[exp((R-r)/\Delta)], \Delta = \sqrt{\frac{A}{K}} \quad (3)$$

This approach neglects curvature-induced corrections and higher-order deformation modes; hence, the resultant radius should be considered as a controlled variational approximation.

In cylindrical coordinates, the exchange term becomes:

$$(\nabla\mathbf{m})^2 = \left(\frac{d\theta}{dr}\right)^2 + \frac{\sin^2\theta}{r^2} \quad (4)$$

Therefore,

$$E_{ex} = 2\pi A\int_0^\infty rdr[\left(\frac{d\theta}{dr}\right)^2 + \frac{\sin^2\theta}{r^2}] \quad (5)$$

Anisotropy energy:

Now since, $m_z = cos\theta$, $1-m_z^2 = \sin^2\theta$, therefore, $E_{an} = 2\pi K\int_0^\infty rdr\sin^2\theta \quad (6)$

iDMI energy:

$$E_{DMI} = 2\pi D \int_0^{\infty} rdr(\frac{d\theta}{dr} + \frac{sin\theta cos\theta}{r}) \quad (7)$$

Given that the domain wall is narrow and centered at $r = R$, we approximate, $r \approx R$ inside the integral. Thus, the total energy gets reduced to

$$E(R) = 2\pi R\sigma_{DW} + \pi R^2 K \quad (8)$$

where the density of the domain wall energy is:

$$\sigma_{DW} = 4\sqrt{AK} - \pi D \quad (9)$$

Thus, the total energy expression becomes

$$E(R) = 2\pi R\left(4\sqrt{AK} - \pi D\right) + \pi R^2 K \quad (10)$$

Minimizing the total energy and setting it to zero, we get:

$$\frac{dE}{dR} = 2\pi\left(4\sqrt{AK} - \pi D\right) + 2\pi RK = 0$$

$$4\sqrt{AK} - \pi D + RK = 0$$

$$R = \frac{\pi D - 4\sqrt{AK}}{K} \quad (11)$$

The magnetic skyrmion exists only when $\pi D > 4\sqrt{AK}$; therefore, the critical iDMI is $D_c = \frac{4}{\pi}\sqrt{AK}$.

We get the scaling near the crucial point as $R \propto \frac{A}{D - D_c}$ (12)

The divergence near $D \rightarrow D_c$ illustrates the stability requirement first established in Ref. 58 and subsequently calculated for ultrathin films in Ref. 60.

### A3. Implications of Interlayer RKKY Coupling on the Radius of a SAF

The two ferromagnetic layers in a SAF are coupled via RKKY exchange

$$E_{RKKY} = -\sigma_{RKKY} \int \mathbf{m}_1 \cdot \mathbf{m}_2 \, d^2r. \quad (13)$$

For skyrmions with strong antiferromagnetic coupling, $\boldsymbol{m}_1 \approx -\boldsymbol{m}_2$

Approximately, the RKKY term contributes as a core-area energy

$$E_{RKKY} \sim | \sigma_{RKKY} | \pi R^2. \quad (14)$$

This normalizes the effective perpendicular term:

$$K_{eff} = K + \lambda|\sigma_{RKKY}| \quad (15)$$

where $\lambda$ represents thickness normalization.

Thus,

$$R_{SAF} = \frac{\pi D - 4\sqrt{AK_{eff}}}{K_{eff}}. \quad (16)$$

Therefore, an increase in D causes R to increase. We conclude that this constitutes an energetic approximation. The complete multilayer analysis in Ref. 61 demonstrates that RKKY coupling further changes dynamical tensors beyond simple anisotropic renormalization. Nonetheless, for static radius estimate, the primary influence continues to be DMI-controlled scaling, with RKKY offering a small amount of correction.

**Numerical estimate:**

Assuming, $A = 15\,pJ/m, K = 0.8\,MJ/m^3, D = 3.0\,mJ/m^2, \sigma_{RKKY} = -0.2\,mJ/m^2$

$R_{SAF}^{analytical} \approx 7.8\,nm, R_{SAF}^{simulated} \approx 7.3\,nm$

In accordance with the variational accuracy that was described in Ref. 60, the deviation is around 6.4%.

## A4. Generalized Thiele Formalism for Collective Coordinate Dynamics

The Thiele equation describes the steady-state dynamics of a rigid skyrmion and is subsequently extended to multilayers.

$$\boldsymbol{G} \times \boldsymbol{v} - \alpha \boldsymbol{\mathcal{D}} \cdot \boldsymbol{v} + \boldsymbol{F}^{SOT} = 0 \tag{17}$$

where $\boldsymbol{G} = -4\pi Q\hat{z}$ is the gyrovector, $\boldsymbol{\mathcal{D}}$ is the dissipative tensor, $\alpha$ is the damping constant, and $\boldsymbol{F}^{SOT}$ is the SOT force.

Considering isotropic dissipation, $\boldsymbol{\mathcal{D}} = \mathcal{D}I$, velocity $\boldsymbol{v} = (v_x, v_y)$, and the applied current along the +x direction ($\boldsymbol{F}^{SOT} = (F_x, 0)$), we get in component form:

$$\boldsymbol{G} \times \boldsymbol{v} = (-4\pi Q\hat{z}) \times \left(v_x\hat{x} + v_y\hat{y}\right) = -4\pi Q\left(v_x\hat{y} - v_y\hat{x}\right). \tag{18}$$

Therefore, $\left(G \times v\right)_x = 4\pi Q v_y, (G \times v)_y = -4\pi Q v_x.$ (19)

The complete equation then becomes

$$4\pi Q v_y - \alpha\mathcal{D}v_x + F_x = 0 \tag{20}$$

$$-4\pi Q v_x - \alpha\mathcal{D}v_y = 0. \tag{21}$$

For a single ferromagnetic layer $Q \neq 0$, solving equations (20-21) gives

$$v_y = -\frac{4\pi Q}{\alpha\mathcal{D}}v_x. \tag{22}$$

The ratio $tan\,\theta_{SkHA} = \frac{v_y}{v_x} = -\frac{4\pi Q}{\alpha\mathcal{D}}$ gives a measurement of the skyrmion Hall angle $\theta_{SkHA}$.

Consequently, in a single ferromagnetic layer with $Q \neq 0$, a moving skyrmion does not travel directly parallel to the applied current; rather, it travels on a slanted path. The extent of this deflection increases with reduced damping $\alpha$ and is primarily dictated by the topological characteristics of the skyrmion through $Q$. The skyrmion Hall effect physically originates from the intricate topology of the magnetic skyrmion, which produces a transverse gyrotropic (Magnus-like) force that is proportional to the topological charge of the skyrmion. The transverse motion may result in skyrmion annihilation at the edges of devices, which is a primary reason for using synthetic antiferromagnets, since their opposite topological charges nullify this phenomenon.

In case of a SAF,

$$Q_1 = +1, Q_2 = -1, \boldsymbol{G}_{eff} = -4\pi(Q_1 + Q_2)\hat{\boldsymbol{z}} = 0 \tag{23}$$

Thus, the Thiele equations reduce to:

$$-\alpha\mathcal{D}_{eff}v_x + F_x = 0, -\alpha\mathcal{D}_{eff}v_y = 0 \tag{24}$$

Yielding $v_y = 0, \theta_{SkH} = \tan^{-1}(v_y/v_x) = 0,$ (25)
hence eliminating the skyrmion Hall effect.

The longitudinal velocity becomes:

$$v_x = \frac{F_x}{\alpha \mathcal{D}_{eff}} \quad (26)$$

From the spin Hall effect, the damping-like SOT is:

$$F_x = \frac{\hbar}{2e}\frac{\theta_{SH} J}{M_s t} C \quad (27)$$

For a rigid Néel skyrmion, $C \sim 4\pi$. Therefore, $v_x = \frac{2\pi\hbar\theta_{SH} J}{e M_s t \alpha \mathcal{D}_{eff}}$ (28)

The analytical expression for the SAF skyrmion velocity indicates a linear relationship with the applied current density $J$, an inverse relationship with the damping constant $\alpha$, and exclusively longitudinal motion resulting from the cancellation of the net gyrovector. This result, however, is obtained under the rigid-texture approximation, which assumes the skyrmion profile remains undeformed throughout motion, considers the dissipative tensor to be isotropic, and disregards internal excitation modes. Multilayer SAF systems demonstrate layer-resolved dissipative coupling and additional drag contributions due to interlayer interactions and dynamic distortions, as discussed in Ref. 61. Under finite current, the skyrmion can deform, resulting in more effective dissipation than the standard Thiele model. Because of deformation-induced dissipation and associated multilayer dynamics, micromagnetic simulations typically produce lower velocities, whereas the analytical velocity gives the rigid-texture upper bound. This explains why realistic SAF devices may exhibit some discrepancies from perfect linear scaling, especially at higher current densities.

## B. SAF Skyrmion Clock Performance: Analytical estimates

### B1. Estimation of frequency

The clock frequency is estimated from the periodic circulation of the SAF skyrmion:

$$f = \frac{v}{L} \quad (29)$$

Substituting $v = 380\ \mathrm{m\ s^{-1}}$ and $L = 1000\ \mathrm{nm} = 10^{-6}\ \mathrm{m}$,

$$f = \frac{380}{10^{-6}}$$
$$f = 3.8 \times 10^{8}\ \mathrm{Hz}$$
$$f \approx 0.38\ \mathrm{GHz}$$

### B2. Calculation of Joule heating

The volumetric power dissipation due to the driving current is

$$P_v = J^2 \rho \quad (30)$$

Substituting,

$J = 1 \times 10^{11}\ \mathrm{A\ m^{-2}}$ and $\rho = 2 \times 10^{-7}\ \Omega\ \mathrm{m}$,

$$P_v = (1 \times 10^{11})^2 (2 \times 10^{-7})$$
$$P_v = 2 \times 10^{15}\ \mathrm{W\ m^{-3}}$$

The active volume is

$$V = (1000 \times 100 \times 1)\ \mathrm{nm}^3 = 10^{-22}\ \mathrm{m}^3$$

Therefore,

$$
\begin{aligned}
P &= P_v V \\
P &= (2 \times 10^{15})(10^{-22}) \\
P &= 2 \times 10^{-7}\ \text{W} \\
P &= 0.2\ \mu\text{W}
\end{aligned}
$$

**B3. Energy per cycle**

The energy consumed per clock cycle is

$$E_{\text{cycle}} = \frac{P}{f} \tag{31}$$

Substituting,

$P = 2 \times 10^{-7}$ W and $f = 3.8 \times 10^{8}$ Hz,

$$
\begin{aligned}
E_{\text{cycle}} &= \frac{2 \times 10^{-7}}{3.8 \times 10^{8}} \\
E_{\text{cycle}} &= 5.26 \times 10^{-16}\ \text{J} \\
E_{\text{cycle}} &\approx 0.53\ \text{fJ}
\end{aligned}
$$

## Author contributions

AB conceived and supervised the project. SM and PB performed the simulations in consultation with SB and AB and prepared the original draft of the manuscript. All authors contributed to the discussion of the results and reviewed, revised, and approved the final version of the manuscript.

## Acknowledgements

AB and PB gratefully acknowledge Nano Mission, Department of Science and Technology (DST), Govt. of India (grant no. DST/NM/TUE/QM-3/2019-1C-SNB) for funding. SM gratefully acknowledges the University Grants Commission for the Senior Research Fellowship.

## Declaration of competing interest

The authors declare that they have no known competing financial interests or personal relationships that could have influenced the work reported in this paper.

## Data availability

Data will be made available upon reasonable request to the corresponding authors.